# Developing and implementing an Einsteinian science curriculum from Years 3 to 10 – Part B: Teacher upskilling: response to training and teacher's classroom experience

Tejinder Kaur[1], Magdalena Kersting[2], Kyla Adams[1], David Blair[1], David Treagust[3], Anastasia Popkova[1], Shon Boublil[1], Jesse Santoso[1], Li Ju[1], Marjan Zadnik[1], David Wood[1], Elaine Horne[1], Darren McGoran[1], Susan Scott[4], and Grady Venville [4]

[1] School of Physics, University of Western Australia, Perth, Australia

[2] Department of Science Education, University of Copenhagen, Copenhagen, Denmark

[3] School of Education, Curtin University, Perth, Australia

[4] Department of Physics, Australian National University, Canberra, Australia

[5] Centre for the Public Awareness of Science, Australian National University, Canberra, Australia

**Abstract**

Recent years have seen a growing interest in modernizing school science curricula to reflect the discoveries in physics since 1900, especially with recent broad recognition of the importance of quantum physics in the modern world. Much effort has been expended in the development of appropriate teaching instruments for teaching Einsteinian physics in schools, but less effort on the crucial topic of teacher professional development. Successful curriculum innovation requires teacher professional development, but for Einsteinian physics we must contend with a lack of confidence due to widely held, but erroneous opinions, that Einsteinian physics is too complex, abstract, and mathematical to be learnt in schools. This paper reports analysis of teacher professional development for practising primary and secondary teachers who were upskilled as part of a process for implementing an 8-year Einsteinian curriculum across 38 primary and secondary schools. Most participants had little prior knowledge of Einsteinian physics. Using self-assessment through questionnaires, interviews, combined with classroom validation, we show that three different professional development programs led to high levels of content knowledge and confidence to teach Einsteinian physics in classes from Year 3 to Year 10. The analysis presented supports our conclusion that it is feasible to upskill teachers from diverse backgrounds in Einsteinian physics and break the cycle that has inhibited the modernisation of school curricula.

## Introduction

### The need for teacher professional development

'Curriculum development must rest on teacher development' (Stenhouse, 1975). This timeless statement, made nearly five decades ago, continues to hold significance in contemporary educational discourse, highlighting the intertwined relationship between curriculum development and teacher professional development. One cannot be successful without the other for several reasons. First, teachers act as the primary implementers of curriculum, and how they are prepared and supported is vital to their success (Jaquith et al., 2010). While the curriculum creates the overarching framework for what should be taught in schools, teachers translate and deliver the curriculum, thereby playing a fundamental role in determining teaching quality and the success of educational systems (Hanushek & Rivkin, 2006; Hattie, 2009; OECD, 2005). Without adequate professional development, teachers may not deliver the curriculum as intended.

Second, teachers need to understand the curriculum content deeply to effectively transform their understanding of it into instruction that their students can comprehend (Shulman, 1986). In addition to curriculum knowledge, Shulman (1986) identified subject-specific and pedagogical content knowledge as key aspects of teachers' professional knowledge. However, in physics, a rapidly progressing discipline, teachers' subject-specific content and pedagogical content knowledge will always need to grow and be updated (Frågåt et al., 2021). Besides, in novel or unfamiliar learning domains, such as Einsteinian physics, teachers' subject-specific content and pedagogical content knowledge is often limited because they were not exposed to the content knowledge during their own education. (Bungum et al., 2015; Farmer, 2021; Frågåt et al., 2022). Without professional development programs focusing on building teacher content knowledge, teachers may struggle to convey the curriculum content and answer students' questions successfully.

Third, curriculum implementation varies based on the educational context and how teachers experience and respond to educational change (Hargreaves, 2005). Different schools, classrooms, and students have different needs, and professional development can help teachers adapt the curriculum to contextual needs (Farmer & Childs, 2022; Hargreaves, 2005). Besides, professional development programmes can provide teachers with opportunities to share their insights, experiences and feedback on the curriculum with professional organizations, education agencies, and policymakers, which, in turn, can inform the curriculum development process (Jaquith et al., 2010). Therefore, professional development provides insights into teachers' perspectives and experiences, which is essential if reform and improvement efforts are to be successful and sustainable (Hargreaves, 2005).

Thus successful curriculum development hinges on sustained teacher professional development. As primary curriculum implementers, teachers require sufficient understanding of the curriculum content, and capacity to adapt it to their unique teaching contexts. Ideally, teachers should play an active role in the curriculum development process.

Professional development becomes particularly significant when introducing novel content, such as Einsteinian physics which represents a complete paradigm shift relative to pre-Einsteinian conceptions of space, time, matter and radiation. Research on professional development in this area is still scant. (Kersting & Blair, 2021).

An approach to equipping teachers for this paradigm shift, and an analysis of teacher responses to professional development forms the basis for this paper. It is the second of two papers on developing and implementing a coherent eight-year Einsteinian science curriculum covering four years of primary school (Years 3-6) and four years of high school (Years 7-10) called Einstein-First. The first paper summarised the curriculum design and student learning outcomes (Kaur et al., 2023).

In this paper, after reviewing previous research in Einsteinian physics professional development, we will present our approach to teacher professional development, and set out three research questions to be answered by this study. We will present the theoretical framework, and methodology which includes the use of questionnaire-based self-assessment. Questionnaires were developed and refined for three professional development formats: one day workshops, in person micro-credential (MC) courses and online MC courses. In total, responses from 37 teachers with or without science backgrounds were assessed. Most of the teachers went on to deliver Einstein-First modules consisting of 8-12 activity-based lessons, defined by detailed lessons plans and instructional videos for teachers. The paper finishes with positive conclusions that support the contention that teachers trained in curriculum-specific skills in the context of clearly defined lessons can effectively deliver Einsteinian physics programs, while all teachers agreed on the importance of modernizing science teaching to reflect the concepts of Einsteinian physics.

### *Previous research on teacher education in Einsteinian physics*

Einsteinian physics plays a significant role in our daily lives and provides the most accurate description of our universe, yet it is not widely taught in science curricula. Existing research indicates that introducing Einsteinian physics into school syllabi poses difficulties for physics teachers because Einsteinian concepts are often perceived as abstract and counterintuitive (Bouchée et al., 2022; Bungum et al., 2015; Frågåt et al., 2022). Teachers find it difficult to understand and impart these concepts to students (Henriksen et al., 2014; Singh, 2008). The reason for the difficulty is that the concepts of relativity, gravity and quantum science represent a radical departure from the classical description of reality. It requires a change in thinking usually only acquired in advanced tertiary-level education, when students are often urged to "forget what you learnt at school". Tertiary courses are typically presented at an abstract mathematical level. Given this level of abstraction, it is unsurprising that teachers perceive Einsteinian physics as quite out of reach and that primary school teachers, particularly those with limited science backgrounds, find it challenging to consider incorporating Einsteinian content into their teaching (Foppoli et al., 2019). Further exacerbating these challenges is the scarcity of well-established instructional strategies tailored to teach these abstract topics (Bungum et al., 2015; Kamphorst & Kersting, 2019).

Professional Development (PD) workshops are essential to enhance teachers' understanding of the subject matter and alleviate their concerns (Phillips, 2008). The professional development required differs from tertiary physics training because it needs to emphasize concepts, offer intuitive understanding, and provide a level of expertise appropriate for school students. Previous research has shown that providing professional learning opportunities can improve teachers' content and pedagogical knowledge (Germuth, 2018). Educators have developed workshops on Einsteinian physics to assess their impact and found that teacher development is an effective way of promoting Einsteinian physics education in high school classrooms (Balta et al., 2019). Another study suggested that delivering a series of workshops and allowing teachers to apply the new knowledge and strategies in the classroom would positively impact students' understanding (Farmer, 2021). Design-based research and close collaboration between physicists, physics educators, and teachers is another approach that has been shown to help teachers grow their subjective-specific content knowledge and pedagogical content knowledge in Einsteinian physics (Frågåt et al., 2022).

### *Research questions*

This paper focuses on teachers' responses to learning and implementing an Einsteinian physics curriculum across school years 3 – 10. We investigate how teachers' subject-specific and pedagogical content knowledge evolves after participating in 75-hour micro-credential courses and one-day professional development workshops. We address the following research questions:

1. What are the most effective strategies for introducing Einsteinian Physics to teachers with limited prior knowledge and experience in this area?
2. How does participation in micro-credential courses and one-day professional development workshops impact teachers' content knowledge, pedagogical expertise and confidence in teaching Einsteinian Physics?
3. Does self-assessment using questionnaires give a consistent measure of the effectiveness of teacher professional development?

The following sections present the methodology, followed by the results, discussion, and conclusion.

## Educational background and theoretical framework

### *Applying the Einstein-First learning approach to teacher training*

The Einstein-First project, described in more detail in Part A, was developed as a collaboration between physicists, science education specialists, and curriculum developers closely associated with a science teacher's association. Activity based learning and the use of models and toys were considered essential to translate the concepts of Einsteinian physics into tangible human-scale form in which students could interact through whole body experiences.

Initially it was assumed that teachers should learn Einsteinian concepts in a more theoretical form than used in the classroom. However, preliminary workshops with the first teachers who implemented Einstein-First programs, showed that the same models and activity-based learning approaches used with students were equally effective in PD programs, and were much easier for teachers to assimilate than a more abstract theoretical approach. Teachers who found theoretical descriptions such as curved space and photons quite baffling, were much more comfortable when the ideas were introduced via toys and models. Models and toys were equally useful for science trained teachers as those without science backgrounds. Most importantly, significant learning was achieved by discussing all the ways that the learning models were *incorrect*. Thus, models support the learning in two ways: one by the usefulness of the analogy that is created, and one by analysing the ways that the analogy is wrong. In all cases the learner carries memory of a tangible object or process to associate with a concept. This is the foundational design concept for our PD program design, and the answer to the first research question. The answer is further elaborated, however, in the context of the evaluation results.

### *The Model of Educational Reconstruction*

Our research approach for this study is underpinned by the Model of Educational Reconstruction (MER), a widely recognized framework for improving instructional practices and teacher professional development programs in science education (Duit et al., 2012). The MER integrates research and development perspectives in three interconnected strands: 1) research on teaching and learning, 2) clarification of content, and 3) evaluation of learning resources. The synergy of the three strands allows us to align our research efforts with effective curriculum development and teacher education strategies. Notably, teacher perspectives played an integral role in each of the three strands of the MER (Figure 1).

After developing the lessons, the team invited primary and high school teachers to informal training sessions. In the sessions, teachers were actively engaged in the activities and asked to identify and select the most effective activities in revealing the key Einsteinian concepts. The lessons were then refined based on this initial teacher feedback and revised again after formal implementation in school trials. The lesson plans and assessment tools developed during this process are available on request.

The teacher training sessions led to further collaboration between teachers and researchers, and these contributed to the PD development in three ways: a) it provided insights into teachers' conceptual development and teachers' confidence for various Einsteinian physics topics; b) it helped clarify content and identify teacher misconceptions, providing a deeper understanding of the material being taught and the areas where clarification was necessary; c) it helped refine learning resources, including lesson plans, learning sequences, activity design and instructional activity videos for teachers. Thus the enhancement of the quality and effectiveness of the Einstein-First curriculum was intimately linked to development of the PD programs.

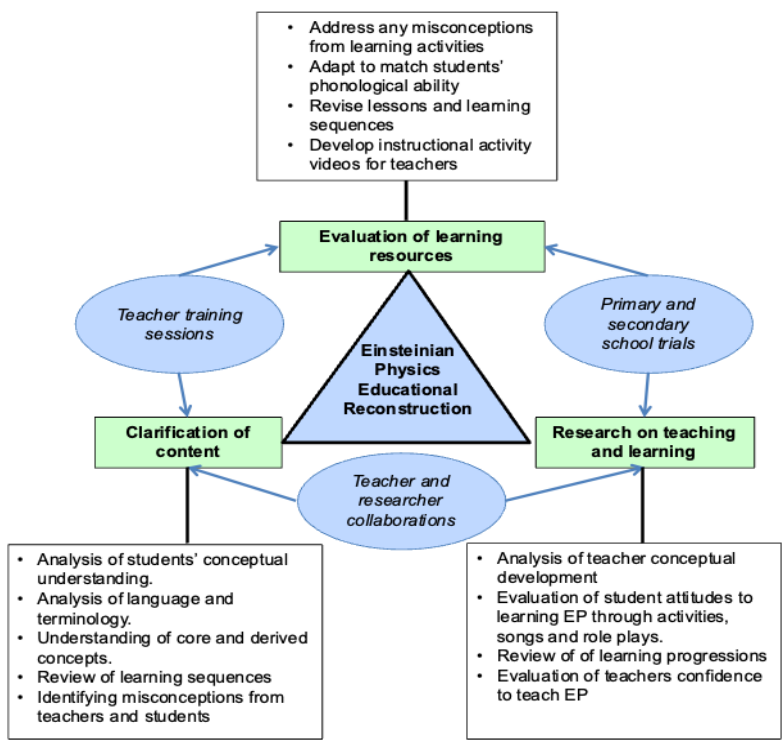

Figure 1: The model of educational reconstruction in which teacher training combines with school trials and collaboration with course designers to optimise the implementation of an Einsteinian science curriculum.

Teacher workshops were designed around Einstein-First lesson plans in line with the spiral-curriculum approach outlined in Part A (Kaur, 2023). As emphasised above, each PD session focused on models and activities, with the concepts elaborated in the context of the activities. For in-person workshops, PD participants undertook group activities following short video illustrations. Online participants participated in live tutorial discussion sessions led by the PD presenter, including video presentation and self-implemented activities when possible.

*Teacher's workshops and micro-credential courses.*

The Einstein-First team developed teacher professional learning workshops (nominally of one day duration) to train teachers in single year topics. As described above, the models and activities used in Einstein-First lessons were used to enhance teachers' content knowledge in Einsteinian physics, and to support them in implementing the program in the classroom.

The content from a large number of one-day workshops were elaborated to create two comprehensive, 75 hour micro-credential courses *Einsteinian physics for primary*

*school teachers* and *Einsteinian physics for secondary school teachers*, each consisting of 37.5 contact hours and 37.5 hours of reading and homework, collectively described as MC1, and first delivered in Jan 2022. Topics included photons, phonons, atoms and molecules, curved space, warped time, gravity, and gravitational waves. Learning was distributed between presentations, activities, self-study and preparation of a presentation to class members. Six presenters who had been involved in creating the Einsteinian physics curriculum delivered the courses.

The courses were repeated in online format in August 2022, described as MC2. Both the primary and secondary courses were delivered in tandem across eight weeks, each involving two hours of pre-recorded lecture-style videos, a one-hour live online tutorial, and five hours of reading, quizzes, and activities that could be done at home. The courses culminated in a two-day hybrid in-person and live online workshop that allowed participants to experience each of the activities interactively. The online course was also assessed by questionnaires and presentations as described above.

After completing the courses, teacher learning was assessed through a live PPT presentation on a proposed lesson chosen by the participant. These were evaluated by two external markers who evaluated based on basic knowledge, confidence and age-appropriateness of the lesson. All participants in MC1 and MC2 were passed on their presentations, thereby providing the team with primary evidence of the adequacy and appropriateness of the PD program, but providing little insight into the participants learning, their difficulties and their confidence.

Questionnaires were chosen to be the primary means for researching the research questions. Different questionnaire formats were chosen for each program to allow the team to simultaneously investigate and optimise the assessment tools (research question 3) while still obtaining data for course evaluation. In principle this approach risked loss of rigour, but for the present exploratory program the observed results had a high degree of consistency, despite the small n-values. The results discussed below, indicate consistency independent of question variations, indicating that the questionnaires can be standardised and simplified for future larger n-value evaluations. They also allowed identification of questions for which there are indications of ambiguous interpretation.

*Participants*

Participants in courses included local, regional and international professional teachers. Data from a total of 37 teachers are included in the data presented here, 22 who participated in professional development workshops and 15 who completed micro-credential courses, as detailed in Table 1.

Table 1: Summary of PD participants reported in this paper. See Part A for module details.

| Teacher Training program | Primary/ Secondary | Questionnaire responses N | Module/s trialled |
|---|---|---|---|
| One-day workshop | Primary | 22 | *Hot Stuff, Atom Frenzy, & Fantastic Photons* |
| Micro-credential course | Primary | 10 | *Hot Stuff, Atom Frenzy, Fantastic Photons & Our Place in the Universe* |
| Micro-credential course | Secondary | 5 | *Warp Spacetime (gravity, year 7), Einsteinian energy (year 8), Quantum world (year 9) and Cosmology (year 10)* |

*Data collection and analysis*

All participants completed questionnaires consisting of both open-ended questions and Likert scale items designed to facilitate and guide self-assessment in relation to three dimensions: a) teacher attitude to the PD program, b) self-assessment of learning relative to the clearly defined learning intentions, and c) teachers confidence to teach Einsteinian physics topics to their school classes. The open-ended questions were marked according to a scale linked to keywords, as described in the results section.

The one-day workshops used 12 Likert-scale questions (Table 2). MC1 used a reduced question set of 9 Likert-scale questions (Table 3) because of observed internal consistency within each dimension. The MC2 course used 11 questions and a 6-point numerical agreement scale (Table 5).

As a means of validating the above assessments, we interviewed a small number of primary school teachers, selected from those who implemented Einstein-First curriculum items in their classrooms after completing the PD program. The interviews aimed particularly to assess teachers' perspectives on lesson plans, identify challenges faced when implementing the program, and gauge their perception of students' reactions and attitudes towards both the activities and the learning topics.

In the next section we will first present an analysis of one day programs, and then analyse the results from MC1 and MC2.

## Results

### *Analysis of Professional Learning*

#### *a) One-day professional learning workshops*

Teachers who participated in the one-day PD workshops were asked to complete a questionnaire and answer a single open-ended question. Table 2 summarises the responses to the twelve questions used to evaluate the three dimensions used for assessment: reaction, learning and confidence. The detailed questions are listed in the table.

Examining the results in each dimension in Table 2, we note first that the responses to the four *reaction* questions was uniformly positive. The uniformity of these responses was the reason that a simplified questionnaire was proposed for the MC1 course evaluation. The learning questions aimed to test four aspects of the learning: first appreciation of a) the need for teaching Einsteinian content, which requires recognition of the new and different physics content, b) recognition that the course provided valuable background knowledge, c) self assessment of learning and d) recognition that the learning matched the learning intentions. The average score for these four question was equal to the first set, but slightly more variable. The third dimension, *teacher confidence*, was slightly weaker due to question 10 which demonstrated that the process for requesting support from the Einstein-First team had not been clearly explained, perhaps also indicating that roughly half the class saw possible need for future support.

Table 2: Questionnaire results conducted with 22 participants who attended one-day professional development workshops. The survey asked participants to rate three dimensions: overall reaction, degree of learning, and confidence in applying what they learned as indicated.

| Item No. | Statement | Number of responses n = 22 | | | |
|---|---|---|---|---|---|
| | **Overall reaction to the workshop** | SD | D | A | SA |
| 1 | I feel positive about this workshop | - | - | 5 | 17 |
| 2 | I found this workshop engaging | - | - | 6 | 16 |
| 3 | The professional learning provided opportunities to collaborate with colleagues | - | - | 5 | 17 |
| 4 | The workshop was well organized | - | - | 5 | 17 |
| | **Degree of learning** | SD | D | A | SA |
| 5 | I appreciate the need for the Einstein-First initiative | - | - | 5 | 17 |
| 6 | The workshop provided the background knowledge needed to plan and teach the Einstein-First module | - | - | 1 | 21 |
| 7 | This professional learning session was a source of learning for me | - | - | 8 | 14 |
| 8 | The learning intentions were largely achieved | - | - | 8 | 14 |

|  | **Degree of confidence to apply learning** | SD | D | A | SA |
|---|---|---|---|---|---|
| 9 | I feel confident that I will be able to integrate the Einstein-First modules into my classes | - | - | 10 | 12 |
| 10 | I am clear what I will need to do to stay connected with the Einstein-First program | - | 1 | 12 | 9 |
| 11 | This professional learning session will help me use the Einstein-First resources to develop my physical sciences program next year | - | - | 6 | 16 |
| 12 | I feel confident to access the online teaching and professional learning | - | - | 5 | 17 |

SD = Strongly Disagree, D = Disagree, A = Agree and SA = Strongly Agree

Overall, the workshop questionnaire results suggest a very high positive reaction, high levels of self-assessed learning and confidence in teaching modern physics concepts. We consider this result quite remarkable considering the fact that most participants had extremely little prior knowledge of Einsteinian physics.

### b) *Micro-credential courses*

The MC course results are analysed separately, because of their different formats and different assessment scales and question structures, used for testing questions and comparison of Likert scale evaluation with a six-point numerical scale, helping to define future assessment approaches for larger numbers of teachers. The high level of consistency between responses to the one-day workshop questionnaires led to a 9-question questionnaire for MC1, which was designed to interrogate the three dimensions of reactions, learning and confidence previously identified. For the online MC2 course that included remote and international participants, we trialled a 6-point numerical scale with 11 questions. The questions and results are shown in Tables 3 and 4 respectively.

Table 3: Questions and results for the face-to-face micro-credential course MC1. Questions 1-3 and 8 focus on overall reaction, 4-6 relate to learning and understanding, while **7** and 9 assess confidence.

| Item No. | Statement | Number of responses n = 8 | | | |
|---|---|---|---|---|---|
|  |  | SD | D | A | SA |
| 1 | I found this course very useful and informative | - | - | 2 | 6 |
| 2 | The course provided opportunities for me to work with other teachers and share my ideas | - | - | 0 | 8 |
| 3 | Participating in the activities and the following discussions were beneficial for my future teaching | - | - | 2 | 6 |
| 4 | I am clear about the resources needed to teach this content | - | - | 2 | 6 |
| 5 | I am clear about the relevance of Einsteinian physics in our daily lives | - | 3 | 1 | 4 |

| 6 | I understand why Einsteinian physics is important to introduce into the school science curriculum | - | - | 2 | 6 |
| 7 | I have hesitations about including Einsteinian physics into my class curriculum | 5 | 1 | 2 | - |
| 8 | The presenters were knowledgeable and prepared for the course | - | - | 2 | 6 |
| 9 | I feel confident presenting an Einstein-First module to my students | - | - | 3 | 5 |

Most of the respondents (6 out of 8) found the course very useful and informative (item 1). At the same time, all participants agreed that the course provided opportunities to work with other teachers and share ideas (item 2).

In terms of the effectiveness of the course activities and discussions, the majority of respondents (6 out of 8) found them to be beneficial for their future teaching (item 3), with all the participants agreeing that they were clear about the resources needed to teach the content (item 4).

The question on relevance in our daily lives requires students to have connected Einsteinian phenomena to everyday observation or technology they use. Only half of the participants strongly agreed that they were clear about relevance, while 3 out of 8 participants expressed disagreement. This concerning result led to greater focus being placed on connecting Einsteinian physics to frequently encountered phenomena. For example, we introduced a training module on quantum spin and its connection to magnetism, added more explicit content on time dilation and GPS navigators, and the connection between quantum probability and partial reflections in windows.

Regarding introducing Einsteinian physics into the school science curriculum, 2 out of 6 participants agreed that they understood why it was important to do so (item 6). In comparison, 5 out of 8 participants felt confident about presenting an Einstein-First module to their students (item 7).

Finally, the participants agreed that the presenters were well-prepared and knowledgeable about the content (item 8), and no participants had any hesitations about including Einsteinian physics in their class curriculum (item 9).

The above questionnaire results can be combined to create a composite class attitude/satisfaction/confidence score by summing and inverting the negative question results. We label the four bands strongly negative, negative, positive, strongly positive. The composite scores from table 3 can be summarised in a histogram, with scores as 0, 5, 15, and 52 as shown in Figure 2.

Three open-ended questions were asked of teachers who completed MC1. They are given in Table 4, including four examples of responses from individual teachers.

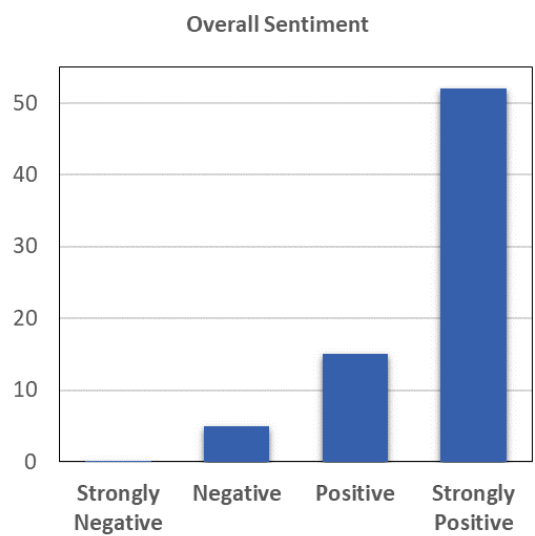

Figure 2: A representation of participants' overall sentiment summarised from Table 4.

Table 4: The open-ended questions and four examples of teachers' responses.

|  | Do you believe that this course has improved your understanding of Einsteinian physics concepts to the point where you feel confident of teaching them in your classrooms? | Did the activities presented in the course meet your expectations? | Did the content of the micro-credential course meet your expectations? |
|---|---|---|---|
| Response 1 | "The lessons were easy and clear to understand, the presenters explained the answers to our questions extensively" | "Yes, age appropriate and engaging" | "Although challenging but I learnt a lot from this course" |
| Response 2 | "Explaining gravity using spacetime was really helpful". | "The activities are simple enough for students to understand" | "Lots of knowledge to take away from this course" |
| Response 3 | "You have taught me clear concepts and provided me with clear lesson plans" | "Great models" | "It provided me the tools and knowledge to teach Einsteinian physics" |
| Response 4 | "Clear teaching background reading material, lots of visual and hands-on activities" | "Activities are easy to set up" | "I do need of many times of reinforces and repetition" |

Table 5: Questions and participant responses from MC2. Numerical scores were requested.

| Item No. | Questions | Number of responses n = 5 | | | | | | Score % |
|---|---|---|---|---|---|---|---|---|
| | | 0 - Not at all | 1 | 2 | 3 | 4 | 5- Completely | |
| 1 | How comfortable would you be to run individual Einstein-First activities within your classroom as part of a trial? | - | - | - | 1 | - | 4 | 92 |
| 2 | How comfortable would you be to run an entire Einstein-First module within your classroom as part of a trial? | - | - | - | 1 | - | 4 | 92 |
| | Going into this course, how important was learning about each of the following aspects of teaching Einsteinian physics? | 0 - Minimally | 1 | 2 | 3 | 4 | 5 – Very important | |
| 3.1 | Conceptual understanding of Einsteinian physics | - | - | - | - | 2 | 3 | 92 |

| | | 0 - Not at all | 1 | 2 | 3 | 4 | 5 - Exhaustively | % |
|---|---|---|---|---|---|---|---|---|
| 3.2 | How to run activities | - | - | - | - | - | 5 | 100 |
| 3.3 | Pedagogical theory | - | - | - | - | 3 | 2 | 88 |
| 3.4 | Implementation of full modules | - | - | - | 1 | - | 4 | 92 |
| To what degree do you feel that the course addressed each of the following aspects? | | 0 - Not at all | 1 | 2 | 3 | 4 | 5 - Exhaustively | - |
| 4.1 | Conceptual understanding of Einsteinian physics | - | - | - | - | 3 | 2 | 88 |
| 4.2 | How to run activities | - | - | - | - | 4 | 1 | 84 |
| 4.3 | Pedagogical theory | - | - | - | 3 | 1 | 1 | 72 |
| 4.4 | Implementation of full modules | - | - | - | 1 | 3 | 1 | 80 |

The MC2 questionnaire allows easy calculation of a percentage score, which is given in the last column. Questions 1 and 2 indicate a high level of confidence to teach the Einstein-First curriculum following the course, with only one participant, who was a pre-service teacher, expressing less than complete confidence in trialling Einstein-First activities or modules within their classrooms.

Questions 3 and 4 are summarised in Figure 3, with Question 3 interrogating participants' expectations and learning priorities going into the course, and Question 4 assessing outcomes of how well each of the learning priorities were addressed by the course.

Question 3 indicates a strong interest amongst participants in all aspects of teaching Einsteinian physics, both conceptual and practical, with the lowest score of 88% being received for *pedagogical theory*.

Questions 4.1, 4.2, and 4.4, scored between 80% and 88%, indicating that teachers were satisfied that the learning they received was near-exhaustive in terms of both conceptual understanding and practical implementation.

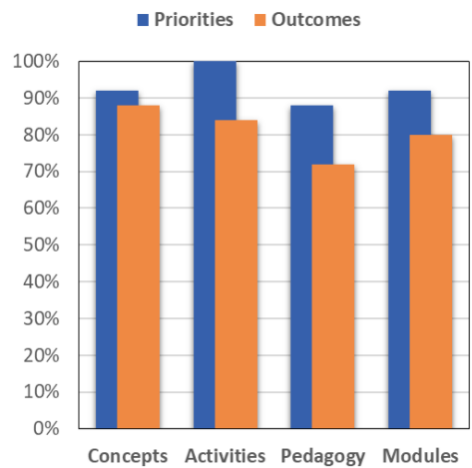

Figure 3: A comparison of participants' learning priorities and learning outcomes, as outlined in Questions 3 and 4 respectively.

Question 4.3 was the only one to score below 80%, which is to be expected as pedagogical theory was not a focus of the course's design. While this is also in line with pedagogical theory being expressed as the participants' lowest relative priority, the strong interest in absolute terms of 88% indicates that future courses may benefit from a more detailed treatment of pedagogical theory as it pertains to the Einstein-First curriculum.

When asked about their experience of attending the activity-focused workshops online, one participant said, "*I attended online, and I do imagine I would have gained more had I been able to come in person, but I was impressed by how much I was able to participate from the online platform. Again, thanks for that!*". A secondary teacher who participated also said, "*Though I already knew quite a bit of the science, I was surprised at how much I learned in this course. The professors gave me new perspectives on subject material I already teach and helped me realize there is a way to teach the Einsteinian concepts in the primary and secondary classroom settings.*"

Based on the data collected, the success of these courses demonstrates that course material focussing primarily on interactive activity-based learning is very effective in teaching the key physical concepts of Einsteinian physics. Teachers can successfully be trained in Einsteinian physics concepts online, allowing us to train teachers across Australia and internationally.

### c) Teachers classroom perspectives: Interviews with selected teachers

Two primary school teachers who completed the face-to-face micro-credential course trialled the Year 3 *Hot Stuff* and *Atom Frenzy* modules, and Year 5 *Fantastic Photons* and *Our Place in the Universe* modules. The team interviewed the teachers about their confidence and experience in delivering the modules. The following paragraphs describe teachers' perspectives regarding the teaching of E-F modules.

After completing the course, a Year 3 teacher described her increased knowledge and confidence as follows: *"The teacher background document was valuable for comprehending the various concepts."* She also attended a one-day workshop and reported that it helped her recall the course material and become more proficient in the activities. The same teacher mentioned that some students were confused by the terminology, despite understanding the concepts through the activities. The terms photons and phonons were confusing for them. She felt that the PowerPoint slides created by the research team were too complex and suggested simplifying them. She also praised the inclusion of YouTube links in the lesson plans, as the students enjoyed learning through videos.

The teacher also mentioned, "*Once students understood atoms and molecules, it was much easier for them to grasp the concepts of solid, liquid, and gas*". During the interview, the teacher stated several times, "*The activities are great*". She also stated, "*Based on my observations, I believe students comprehend the concepts. However, they appear confused about the terminology*" In response to this interview songs were designed to help develop students' science vocabulary and address phonological problems as discussed in part A. The songs are available on youtube (https://www.youtube.com/watch?v=AxSgElLe9Ig)

The same teacher expressed difficulty organizing equipment and requested boxes containing all the necessary materials. This request has since been implemented. She concluded by stating, "*After completing the micro-credential course, I had gained the knowledge and confidence to deliver the program in the classroom, and the one-day workshops further boosted my confidence. A few concepts were unclear to me during the course, but I now have a firm grasp on them*".

A year five teacher attended a one-day training workshop before the micro-credential course had trialled the *Fantastic Photons* module. During the course, she noted that "*many of the concepts were not clear to me when I was delivering the program. However, by attending this course, I am gaining the knowledge I lacked during my trials.*" After completing the micro-credential course, the teacher taught two Year 5 modules. The teacher also taught chemistry and introduced atoms and molecules using Year 3 Einstein-First lesson plans.

In response to a question about confidence after the MC1 course, she stated, "*I am much more confident now, and occasional one-day workshops are very helpful.*" In addition, she said "*Now that I have completed the micro-credential course, I am able to answer students' questions.*"

When asked if students were confused about the terminology, the teacher stated, "*There was no confusion regarding terminology; I made a few things clear at the beginning of the program, such as bulletiness is a made-up word to emphasize the particle nature of light.*" She also stated that some PowerPoint slides should be simplified.

It is important to note this teacher also indicated certain difficulties. She found it challenging to complete all the lessons. She requested additional information in background information about photons and also requested boxed equipment for each module and suggested that certain YouTube videos were too advanced.

The above is an example of how the Einstein-First curriculum has been created in close collaboration with teachers with the PD component integral to the overall development of the program. We go on to discuss the results in the context of the research questions.

### Discussion

The primary objective of this study was to determine whether teacher professional development can effectively upskill teachers with little background in physics and no prior knowledge of Einsteinian physics, to the extent that they are able to develop sufficient content knowledge to teach activity-based lessons in Einsteinian physics.

Research question 1 asked about the most effective strategies for introducing Einsteinian Physics to teachers with limited prior knowledge. Through initial trials we had identified the strategy that we went on to trial in the three PD programs reported here. These confirmed the effectiveness of upskilling based on the activities and models that had already been proven effective for teaching students. The very high value placed on activities by the teacher/participants in the PD courses reported above support the conclusion that learning with physical models is more effective than approaches that rely on documents or videos alone. The findings resonate with previous studies suggesting that the choice of teaching methods plays a crucial role in enabling teachers to assimilate and later convey complex concepts to their students (Darling-Hammond et al., 2017; Farmer, 2021).

Research question 2 asked how teacher PD impact teachers' content knowledge, pedagogical expertise and confidence. Again, the questionnaire results, combined with interview results, open ended questions, and successful live presentations provide strong evidence that the PD approach used has effectively enabled teachers to transition to an Einsteinian model of reality, and to be able to present it with confidence, at least within the confines of a clearly defined curriculum.

The successful implementation of Einstein-First programs in the classroom by some participants supports the view that the teacher PD was crucial for all the components of question 2. This aligns with research suggesting that intensive, content-focused professional development can improve teaching practice and student outcomes (Darling-Hammond et al., 2017; Desimone, 2009). Notably, our study intersects with findings from Farmer and Child's 2022 study: our results underline the value of networking and collaboration among teachers, with participating teachers expressing a desire for more time for collaborative planning and resource development based on easily accessible teaching material. Farmer's research also highlighted the value of collaboration in teacher professional development, finding that rural Scottish teachers highly valued learning from their departmental colleagues, formal and informal interactions, and sharing good practices (Farmer & Childs, 2022).

Our research results, presented in Tables 2, 3, 4 and 5, demonstrate how teachers from diverse backgrounds trained in Einsteinian physics effectively delivered the program and improved students' conceptual understanding of Einsteinian physics (Kaur et al., 2023). These findings are consistent with previous research in which Einsteinian physics lessons were delivered by expert educators and researchers, who observed a significant increase in student comprehension after the program (Kaur et al. 2018. Kaur et al. 2019. Choudhary et al 2019. Foppoli et al 2020, Adams et al 2021. Kersting et al 2018., Matteo et al. 2022).

Research question 3 asked about our primary means of collecting data using self-assessment via questionnaires. We have already noted the remarkable self-consistency of responses, but also the means of validating their responses through interviews and open ended questions. The results from numerical agreement values as opposed to agree/strongly agree options were also consistent except for internal inconsistencies remarked on in the previous section, thought to be due to poor question wording. In all forms of testing, teachers highly valued hands-on activities. They felt these activities facilitated their understanding and assisted them in introducing students to the concepts of Einsteinian physics.

The teachers' experiences are consistent with the broader research literature indicating that active learning approaches effectively increase student engagement and learning outcomes in science education (Hodson, 2014; Hofstein, 2017; Kersting et al., 2023; Lunetta et al., 2007). A comment from a science coordinator where the program was trialled with year 3, year 4 and Year 5 classes underscores the success of the hands-on approach in teaching and learning Einsteinian physics: *"The Einstein-First Project provides improved access to quality STEM education through the development of curriculum and resources. In addition, by providing teachers with professional development it allows these concepts to be taught. Across our student cohorts we have a wide range of abilities; from gifted and talented to ESL (English as a second language) and learning or social disabilities such as autism, ADHD and dyslexia for example. A principal strength of the units of work is the 'hands on' physical representations of physics concepts. The team have translated quite complex concepts into elegant activities that the students can see and interact with, which teachers can use to teach. For example, the spacetime simulator is a highly engaging kinetic and visual learning experience which is accessible to students of all abilities".*

Further insights in how active learning of Einsteinian physics improved the teaching experience are given by a year 8 teacher: *"The notable thing about the Einsteinian physics lessons is that students are fully engaged, disruption is rare, and the students with learning difficulties are practically indistinguishable from the mainstream students. The classes are easy to teach because the students are so engaged".*

A year 5 teacher commented that *"The activities and experiments included in the program provide rich learning opportunities to explore important and foundational theories about our universe. I saw*

*remarkable gains in achievement and understanding when comparing the pre and post-assessment results of my students. I would recommend this program to any teachers that are interested in providing an engaging, accurate and relevant science curriculum for their students".*

Finally, the application of the Model of Educational Reconstruction to facilitate teacher professional development in this study highlights the effectiveness of this model in aligning the training received by educators with the curriculum content and teaching strategies. This confirms the previous research by Duit et al. (2012) that emphasized the importance of empirical research in shaping curriculum development and teacher training programs.

While this study provides insights into the efficacy of our professional development workshops and micro-credential courses in Einsteinian physics, it is crucial to acknowledge certain limitations associated with the research design.

We have taken a pragmatic approach that assumes it is impossible to expect most teachers to have advanced qualification in physics. Our claim is that it is not necessary to have advanced qualifications to understand physical reality at the level of the underlying concepts. Quite clearly, our pragmatic approach is not designed to develop mastery of physics, but it is designed to create an appetite for deeper understanding.

One-day workshops only provide teachers with a narrow range of content. Several teachers in our cohort subsequently came to the MC courses because they recognised the lack of breadth in their workshop training. The MC courses might be considered as not offering sufficient time or opportunities for teachers to engage in reflective practice. We suggest that this limitation is overcome through teacher's increased confidence and repeated implementation in the classroom.

We look forward to a future PD environment in which a combination of training approaches, including longer-term professional development programs and follow-up support, will allow teachers' to steadily increase their knowledge and improve their pedagogical practices.

This study had a small sample size, with data from 22 teachers participating in one-day workshops and 15 teachers completing the MC courses. As the Einstein-First project expands to more schools across Australia, we anticipate much more data and more reliable results.

We were worried that the self-assessment approach using questionnaires would lead to unreliable data influenced by social desirability bias (Over-emphasis on positive experiences). However the level of consistency between self-assessments across three different questionnaire formats, combined with the quality of their PPT presentations, open-ended questions and interviews led us to be confident of the validity of the assessments, especially when they are matched with the learning outcomes reported in part A. Further research could consider incorporating classroom observations to assess the implementation of the Einsteinian physics curriculum more holistically.

Despite the above limitations, we believe that our study provides an important demonstration of the effectiveness of professional development in Einsteinian physics education.

**Conclusion**

The importance of teacher professional development in implementing innovative curriculum initiatives has been reaffirmed by this study, in the context of the novel area for school science of Einsteinian physics that involves a substantial change of paradigm. Our findings are in line with the assertion by Jaquith et al. (2010) that professional development is the most critical strategy for extending and refining teachers' knowledge, skills, and practices throughout their careers. It provides confidence that teacher professional development can be effectively linked to the latest discoveries in science. (Frågåt et al., 2021).

In part A we presented student learning outcomes from our Einstein-First program (Kaur et al., 2023). In this paper to have contrasted three teacher PD programs, and shown that all three enabled the widespread implementation of a curriculum based on our current best understanding of physics. This study has shown that teacher training focused on the essential concepts of Einsteinian physics, and taught through physical activities, improves teachers' confidence in teaching this unfamiliar subject in their classrooms.

All the participants in this study reported confidence in their ability to explain key concepts and engage their students in activities that promote a deeper understanding of our universe, while teachers who had trialled programs in their classrooms reported numerous benefits. The three methods trialled in this study, specialised workshops, or comprehensive face-to-face and online micro-credential courses all enabled teachers with little science training to teach modern physics content. Thus, our project represents an achievable method for upskilling teachers on a large scale so that it becomes feasible to modernize science education across all schools without the need for massive levels of teacher re-training.

In conclusion, this study provides valuable insights into how teachers can be equipped to introduce Einsteinian physics into their classrooms. Our findings indicate that professional development is critical in successfully implementing an ambitious curriculum change. Through workshops and micro-credential courses, teachers can gain the necessary knowledge and confidence to teach Einsteinian physics effectively, even with limited prior experience in this area.

We showed how the Model of Educational Reconstruction linked curriculum development to professional development. It was instrumental in ensuring alignment between teacher training, curriculum content, and teaching strategies. The overwhelmingly positive response from teachers and the perceived improvement in their teaching practice underscores the feasibility of this approach.

In past decades the modernization of science education through introduction of Einsteinian physics as the central pillar for understanding of the world around us, was generally thought to be beyond the abilities of children. Part A in this series demonstrated children's learning across ages 7-15, while this paper demonstrates an effective approach to teacher professional development. There is an evident need to take this research further, because our sample sizes were small. However, we believe we have shown how to break the cycle that has inhibited the modernization of school curricula to date.

Our findings lend weight to the argument for an Einstein-First approach to physics education, demonstrating its potential to transform school science, ensuring that *all* students' are able to share the revelations of modern science. The consequent improvement in national scientific literacy should enable them as adults to participate in public discourse, develop informed opinions on science and technology issues, as well as encourage them to pursue careers in science and technology.

## Acknowledgments

We would like especially to thank two volunteers on the Einstein-First team, Peter Rossdeutscher and Howard Golden, who have enabled us to raise donation funds to supplement our ARC Linkage funding (LP180100859) that allowed us to develop our on-line training programs. We also thank the ARC Centre of Excellence for Gravitational-Wave Discovery (OzGrav) for their continual support, especially in enabling us the develop school kits for our activities. We also thank our Einsteinian Physics Education Research (EPER) collaborators for their enthusiastic support. We are very grateful to the West Australian Department of Education for their support, the Independent Schools Association of Western Australia who have facilitated many of our trials, and the Science Teachers Association of Western Australia who have provided essential and continuous support. We are grateful to our participating schools principals, teachers, and students for allowing us to run the program and for granting us permission to use their photographs and data for research purposes. Our program is supported by numerous companies listed on our website www.einsteinianphysics.com.

## Declaration of interest statement

The authors declare that they have no conflicts of interest related to this research.

## Ethics statement

All of the participating schools' principals, teachers, students, and parents provided their informed consent. In addition, students were informed that participation in the Einstein-First project was voluntary and would not affect their school grades. The research design followed ethical standards-compliant with the Research Ethics Committee at the University of Western Australia, the department of Education Western Australia, Association of Independent schools of Western Australia and the Catholic Education Western Australia. The data is stored securely on the Institutional Research Data Store (IRDS) system at the University of Western Australia.